\documentclass[reprint,amsmath,twocolumn,superscriptaddress]{revtex4}

\usepackage{amsmath,amssymb,graphicx}
\usepackage{ mathrsfs }
\usepackage{xcolor}
\usepackage{ulem}

\newcommand{\ket}[1]{\left|#1\right\rangle}
\newcommand{\bra}[1]{\left\langle#1\right|}
\newcommand{\braket}[2]{\left\langle#1|#2\right\rangle}

\newcommand{\comm}[2]{\left[#1,#2\right]}

\newcommand{\der}[2]{\frac{d\, #1}{d\, #2}}

\newcommand{\pder}[2]{\frac{\partial\, #1}{\partial\, #2}}

\newcommand{\cc}{c.c.}

\begin{document}
\title{Optimal Protocols in Quantum Annealing and QAOA Problems}
\author{Lucas~T.~Brady}
\email{Lucas.Brady@nist.gov}
\affiliation{Joint Center for Quantum Information and Computer Science, NIST/University of Maryland, College Park, Maryland 20742, USA}
\affiliation{Joint Quantum Institute, NIST/University of Maryland, College Park, Maryland 20742, USA}

\author{Christopher~L.~Baldwin}
\affiliation{Joint Center for Quantum Information and Computer Science, NIST/University of Maryland, College Park, Maryland 20742, USA}
\affiliation{Joint Quantum Institute, NIST/University of Maryland, College Park, Maryland 20742, USA}

\author{Aniruddha~Bapat}
\affiliation{Joint Center for Quantum Information and Computer Science, NIST/University of Maryland, College Park, Maryland 20742, USA}
\affiliation{Joint Quantum Institute, NIST/University of Maryland, College Park, Maryland 20742, USA}

\author{Yaroslav~Kharkov}
\affiliation{Joint Center for Quantum Information and Computer Science, NIST/University of Maryland, College Park, Maryland 20742, USA}
\affiliation{Joint Quantum Institute, NIST/University of Maryland, College Park, Maryland 20742, USA}

\author{Alexey~V.~Gorshkov}
\affiliation{Joint Center for Quantum Information and Computer Science, NIST/University of Maryland, College Park, Maryland 20742, USA}
\affiliation{Joint Quantum Institute, NIST/University of Maryland, College Park, Maryland 20742, USA}

\date{\today}
\begin{abstract}
Quantum Annealing (QA) and the Quantum Approximate Optimization Algorithm (QAOA) are two special cases of the following control problem: apply a combination of two Hamiltonians to minimize the energy of a quantum state.
Which is more effective has remained unclear.
Here we analytically apply the framework of optimal control theory to show that generically, given a fixed amount of time, the optimal procedure has the pulsed (or ``bang-bang") structure of QAOA at the beginning and end but can have a smooth annealing structure in between.
This is in contrast to previous works which have suggested that bang-bang (i.e., QAOA) protocols are ideal.
To support this theoretical work, we carry out simulations of various transverse field Ising models, demonstrating that bang-anneal-bang protocols are more common.
The general features identified here provide guideposts for the nascent experimental implementations of quantum optimization algorithms.
\end{abstract}

\maketitle

\textit{Introduction.}
The ongoing development of Noisy Intermediate Scale Quantum devices is guided by the question of how to leverage limited resources to best prepare the desired state of a system.
Both Quantum Annealing (QA)~\cite{Kadowaki,Farhi2000} and the Quantum Approximate Optimization Algorithm (QAOA)~\cite{Farhi2014} search for a combination of two Hamiltonians which prepares the ground state of one of the Hamiltonians as quickly and accurately as possible. 
QA smoothly interpolates between the two Hamiltonians, whereas QAOA applies one or the other in sequence.
It has previously been unclear which method, if either, is the most efficient.

Previous works~\cite{Yang,Bapat2018,Mbeng,Lin} have applied the formalism of optimal control theory, in particular Pontryagin's Maximum/Minimum Principle~\cite{Pontryagin}, to this problem.
It has been suggested on the basis of Pontryagin's principle that a ``bang-bang'' protocol, as in QAOA, is optimal~\cite{Yang}.
Yet as we  demonstrate, some of the assumptions behind this result are not true in general.
We show that hybrid protocols, consisting of both bang-bang and annealing segments, are often best, backing up our analytic results with numerics for Ising models.
Indeed, Ref.~\cite{Lin} found that a bang-anneal-bang protocol is optimal for the unstructured search problem.
Our work elucidates this observation, and extends it further.

QA is closely related to Adiabatic Quantum Computing~\cite{Farhi2000}, in which the Hamiltonian interpolates from a simple ``mixer'' to the desired ``problem'' Hamiltonian.
The adiabatic theorem guarantees that the system, if initially in the ground state of the mixer Hamiltonian and deformed sufficiently slowly, will remain in the ground state throughout.
QA generalizes this to allow for non-adiabatic protocols.
Even in adiabatic regimes, optimization of the annealing schedule can potentially give polynomial speedups over both classical algorithms and unoptimized quantum schedules~\cite{Roland,Jarret}.

QAOA, on the other hand, applies the mixer and problem Hamiltonians alternately, using the timings of these pulses (``bangs'') as variational parameters to be optimized over~\cite{Farhi2014}.
There is evidence that restricted forms of QAOA are more powerful than \textit{adiabatic} protocols with the same restrictions~\cite{Farhi2016} (although both are known to be quantum universal~\cite{Aharonov,Lloyd}), but this does not address whether a non-adiabatic annealing procedure can outperform QAOA.

Optimal control theory~\cite{Pontryagin} is well-suited to address such questions.
It has long been used in a variety of physics and chemistry fields~\cite{Rabitz2000,Werschnik,Brif11,Palao03, Palao02, Cui, Omran, Montangero, Gorshkov, Khaneja, Grace}.
Applications to QA/QAOA are more recent, beginning with Ref.~\cite{Yang} and continuing with Refs.~\cite{Bapat2018,Lin,Mbeng}.
The QA/QAOA problem is distinct from the majority of quantum optimal control problems in that it has only one control function to be applied in a large Hilbert space.
As a result, unlike in standard quantum optimal control, the desired state typically cannot be prepared exactly in finite time.
Some work has been done to examine the standard theory in this limit~\cite{Riviello15}, but the results in this direction remain sparse.

In what follows, we first carefully articulate the control problem under consideration, pulling in standard ideas from optimal control and applying them to the QAOA/Annealing problem. We then prove general statements about where bang-bang vs.\ annealing forms are preferred and what forms those will take in practice and finally present numerical results that support the analytic results.  Our results also provide both a novel recasting of QAOA within the Quantum Annealing framework and an analysis of Quantum Annealing in the low time regime where it deviates heavily from the adiabatic dynamics that inspired it.

\textit{Control Problem.}
The problem which both QA and QAOA seek to solve is as follows: given Hamiltonians $\hat{B}$ and $\hat{C}$, with the system in the ground state of $\hat{B}$ at time 0, find the protocol $u(t)$ which minimizes the energy
\begin{equation} \label{eq:final_energy_expression}
J \equiv \langle x(t_f) | \hat{C} | x(t_f) \rangle ,
\end{equation}
where the time evolution of $| x(t) \rangle$ is given by
\begin{equation} \label{eq:Schrodinger_equation}
\begin{gathered}
\frac{\textrm{d}}{\textrm{d}t} | x(t) \rangle = -i \hat{H}(t) | x(t) \rangle , \\
\hat{H}(t) \equiv u(t) \hat{B} + (1 - u(t)) \hat{C} .
\end{gathered}
\end{equation}
To avoid extreme protocols, we require that
\begin{equation} \label{eq:protocol_limits}
u(t) \in [0, 1] \; \; \forall t.
\end{equation}
Our analytic results apply for any two Hamiltonians where the initial state is the ground state (or any eigenstate) of $\hat{B}$ and the target state is the ground state of $\hat{C}$.  Often in Annealing/QAOA problems $\hat{C}$ (the problem Hamiltonian) is diagonal in the computational basis, while $\hat{B}$ (the mixer Hamiltonian) is a transverse field on the qubits.
As stated above, QA assumes a smooth $u(t)$ whereas QAOA assumes that $u(t)$ jumps suddenly between $0$ and $1$ in bangs.

To apply optimal control theory to this problem, we interpret Eq.~\eqref{eq:Schrodinger_equation} as a constraint relating $| x(t) \rangle$ to $u(t)$ and account for it by introducing a Lagrange multiplier $| k(t) \rangle$.
Thus the cost function is modified to
\begin{equation} \label{eq:control_action}
\begin{aligned}
J =& \; \langle x(t_f) | \hat{C} | x(t_f) \rangle \\
+& \int_0^{t_f} \textrm{d}t \, \langle k(t) | \left[ -\frac{\textrm{d}}{\textrm{d}t} - i \hat{H}(t) \right] | x(t) \rangle + \textrm{c.c.}.
\end{aligned}
\end{equation}
Optimal control theory~\cite{Pontryagin, Kirk} then uses calculus-of-variations techniques to derive necessary conditions for a minimum of $J$ (see Appendix \ref{app:control} for the full derivation of these equations):
\begin{equation} \label{eq:optimal_x_equation}
\frac{\textrm{d}}{\textrm{d}t} | x(t) \rangle = -i \hat{H}(t) | x(t) \rangle ,
\end{equation}
\begin{equation} \label{eq:optimal_k_equation}
\frac{\textrm{d}}{\textrm{d}t} | k(t) \rangle = -i \hat{H}(t) | k(t) \rangle ,
\end{equation}
\begin{equation} \label{eq:optimal_endpoint_equation}
| k(t_f) \rangle = \hat{C} | x(t_f) \rangle ,
\end{equation}
and finally, for all \textit{allowed} variations $\delta u(t)$ of the protocol, 
\begin{equation} \label{eq:optimal_u_equation}
\begin{aligned}
&\frac{\delta J}{\delta u(t)} \delta u(t) \equiv  \; \Phi(t) \delta u(t) \geq 0, \\
&\Phi(t) = \left[ i \langle k(t) | \big( \hat{C} - \hat{B} \big) | x(t) \rangle + \textrm{c.c.} \right].
\end{aligned}
\end{equation}

Note that Eq.~\eqref{eq:optimal_u_equation}, which is a form of Pontryagin's Minimum Principle \cite{Pontryagin} applied to this problem, can be satisfied at any given time in one of three ways: i) $\Phi(t) = 0$; ii) $\Phi(t) > 0$ and $u(t) = 0$; iii) $\Phi(t) < 0$ and $u(t) = 1$.
The first possibility, that the functional derivative is 0 at the minimum of $J(t)$, is natural from a calculus-of-variations perspective.
The latter two are legitimate only because $u(t)$ is restricted to be between 0 and 1, and Eq.~\eqref{eq:optimal_u_equation} needs to hold merely for all \textit{allowed} $\delta u(t)$.
However, situations in which $\Phi(t) = 0$ for an extended interval have historically been referred to as ``singular''~\endnote{The terminology is due to the fact that, for problems linear in both the state $x(t)$ and control $u(t)$ without their product (unlike the Schrodinger equation), one can show that $\Phi(t)$ can only vanish in a finite-length interval if the control matrices are non-invertible.}.
Previous works have argued that such situations are uncommon in practice and thus that the optimal protocol must be of bang-bang form~\cite{Yang}.
One of our key results is that singular regions are in fact quite natural, meaning that the exceptions noted in Ref.~\cite{Yang} are often the rule.
This point is obscured by the fact that some common classical systems cannot exhibit such singular regions and are always bang-bang \cite{Pontryagin,Kirk}.

Through Eq.~\eqref{eq:optimal_u_equation}, we learn the following about the QA/QAOA problem: if an optimal protocol has a smooth annealing form in some interval, then $\Phi(t)$ must equal zero in that interval.
Correspondingly, if $\Phi(t)$ is non-zero in some interval, then the protocol must be of bang-bang form in that interval.

Just as one transforms the Lagrangian of a dynamical system into a Hamiltonian by the Legendre transform, one can construct from the cost function $J$ a ``control Hamiltonian'' $\mathbb{H}$ (not to be confused with the system Hamiltonian $\hat{H}$).  For our problem, the control Hamiltonian evaluates to
\begin{equation} \label{eq:control_Hamiltonian}
\mathbb{H}(t) = i \bra{k(t)}\hat{H}(t)\ket{x(t)} + \textrm{c.c.}.
\end{equation}
The derivation of $\mathbb{H}$, as well as a proof that it is a constant in time, is carried out in Appendix \ref{app:con_ham}.

Eq.~\eqref{eq:optimal_u_equation} is often expressed in terms of the control Hamiltonian, in which case singular regions are defined as those where $\delta \mathbb{H} / \delta u(t) = 0$. This is equivalent to the definition used here, since (see Eq.~\eqref{eq:control_Hamiltonian_alternate} below) $\delta \mathbb{H} / \delta u(t) = -\Phi(t)$.

\textit{Time constraints.}
It is important
that we restrict to a fixed runtime $t_f$. 
If protocols are allowed arbitrarily long times, then the problem becomes trivial: the adiabatic theorem guarantees that any sufficiently slow protocol will end in the desired ground state.
Since adiabatic protocols are often prohibitively inefficient, we constrain ourselves to more feasible runtimes.
One way to do so is to simply fix $t_f$ (``hard constraint''), as we have done.
Another would be to allow protocols of varying $t_f$ but include a penalty term $\lambda t_f$ (``soft constraint'') in the action [Eq.~\eqref{eq:control_action}].

There is a useful connection between these two means of enforcing the time constraint which shows that they ultimately yield the same protocol.
Furthermore, this connection gives physical meaning to the control Hamiltonian $\mathbb{H}$.  The value of $\mathbb{H}$ in a hard-constraint problem with given $t_f$ equals the value of $\lambda$ needed in a soft-constraint problem for the optimal protocol to have the same runtime $t_f$.
We prove this equivalence in detail in Appendix \ref{app:soft_con}.

As a final point, we will assume that $t_f$ is small enough that the desired ground state cannot be reached exactly, which is often the case in the setting of variational state preparation.
Given enough time, the control problem is under-constrained and has several optimal solutions that may in general be hard to characterize.
This assumption implies that $\mathbb{H}$ is strictly positive, since some amount of penalty is needed for $t_f$ to be the optimal runtime.

\textit{Initial and final bangs.}
We now show that any optimal protocol for our control problem must
both begin and end with a bang.
For some finite time interval at the beginning, the protocol must have $u(t) = 0$, and for another finite time interval at the end, it must have $u(t) = 1$.

To prove this, write $\Phi(t) = \Phi_C(t) - \Phi_B(t)$, where
\begin{equation} \label{eq:phi_B_definition}
\Phi_X(t) \equiv i \langle k(t) | \hat{X} | x(t) \rangle + \textrm{c.c},
\end{equation}
for any operator $\hat{X}$.
Note that $\Phi_B$, $\Phi_C$, and thus $\Phi$ are continuous functions of time, as is clear from the continuity of $\ket{x}$ and $\ket{k}$ [see Eqs.~\eqref{eq:optimal_x_equation} and~\eqref{eq:optimal_k_equation}].
Also, the control Hamiltonian can be written as
\begin{equation} \label{eq:control_Hamiltonian_alternate}
\mathbb{H} = u(t) \Phi_B(t) + (1 - u(t)) \Phi_C(t).
\end{equation}

Consider the final portion of the protocol first.
Eq.~\eqref{eq:optimal_endpoint_equation} gives $\Phi_C(t_f) = \textrm{Re}[i \langle x | \hat{C}^2 | x \rangle ] = 0$.
Eq.~\eqref{eq:control_Hamiltonian_alternate} then gives $\Phi_B(t_f) > 0$ (remember that $\mathbb{H} > 0$), and thus $\Phi(t_f) < 0$.
The continuity of $\Phi(t)$ then implies that $\Phi(t) < 0$ for a finite interval ending at $t_f$.
We thus have that $u(t) = 1$ for a finite interval at the end of the protocol.

The initial portion of the protocol can be treated similarly, albeit with one additional step.
Note that by Eqs.~\eqref{eq:optimal_x_equation} and~\eqref{eq:optimal_k_equation}, the time derivative of $\langle k(t) | x(t) \rangle$ is 0, thus $\langle k(0) | x(0) \rangle = \langle k(t_f) | x(t_f) \rangle$.
Since the system is assumed to initially be in the ground state of $\hat{B}$,
\begin{equation} \label{eq:initial_bang_derivation}
\begin{aligned}
\Phi_B(0) =& \; \textrm{Re} \big[ i \langle k(0) | \hat{B} | x(0) \big] \\
\propto & \; \textrm{Re} \big[ i \langle k(0) | x(0) \rangle \big] = \textrm{Re} \big[ i \langle k(t_f) | x(t_f) \rangle \big] .
\end{aligned}
\end{equation}
Eq.~\eqref{eq:optimal_endpoint_equation} thus gives $\Phi_B(0) = 0$.
Identical reasoning to above then shows that $\Phi(t) > 0$ and $u(t) = 0$ for a finite interval at the beginning of the protocol.

These results make sense heuristically.
At the beginning of the protocol, the system is in an eigenstate of $\hat{B}$ and thus application of $\hat{B}$ does nothing to the state.
While the final state is not exactly an eigenstate of $\hat{C}$, it should be close to the target state which is the ground state of $\hat{C}$.  Therefore, roughly we might expect that similar logic about the futility of $\hat{C}$ at the end might hold.
Our results above prove that this is indeed the case.

In Appendix \ref{app:bang_len}, we discuss how the lengths of the initial and final bangs vary with the parameters of the problem.
In particular, we find that they become small as $t_f$ increases, with the procedure approaching the monotonic annealing schedule typical of adiabatic quantum computing.  Our results can be interpreted as saying that optimal quantum annealing for short timescales deviates from the monotonic ramp characteristic of the adiabatic regime.

The condition for a singular region is that $\Phi(t)$ and all its time derivatives must be zero for a finite interval of time.  This condition can be achieved in a variety of ways.  It is known in classical systems \cite{Kirk} that uncontrollability can lead to singularities, but our system is guaranteed to be able to reach the desired state due to the adiabatic theorem since the initial and final states are the ground states of the two Hamiltonians.  Even without reachability problems, there are other forms of singularities that exist and impose conditions on the optimal $u(t)$.  In Appendix \ref{app:singularities}, we explain these conditions.  Numerous options are possible, largely based on the structure of the control space and how the derivatives of $\Phi(t)$ are set to zero.  We put particular emphasis on the simplest to derive condition which is the form we have seen for all singular regions numerically and is given by:
\begin{equation}
     \label{eq:u_sing}
     u^*(t) = \frac{\Phi_{\comm{\comm{\hat{B}}{\hat{C}}}{\hat{C}}}(t)}{\Phi_{\comm{\comm{\hat{B}}{\hat{C}}}{\hat{B}}}(t)-\Phi_{\comm{\comm{\hat{B}}{\hat{C}}}{\hat{C}}}(t)}.
\end{equation}
While this equation describes how a singular region behaves, it does not specify whether the critical point/minimum determined by such a singular condition will be the global versus local minimum, for which we turn to numerics.

\textit{Numerical results.}
The basic form and structure of the optimal protocol has been discussed for general Hamiltonians.  The variety of singularity types and bang-bang possibilities in the middle region turns out to be problem dependent.  To shed light on these concepts in practice, we here support our analytic results with extensive numerical simulations for transverse-field Ising models; though, our analytic results are valid for other models.
The mixer and problem Hamiltonians are
\begin{equation} \label{eq:TFIM_transverse_field}
\hat{B} = -\sum_{i=1}^N \hat{\sigma}_i^x,
\end{equation}
\begin{equation} \label{eq:TFIM_interactions}
\hat{C} = \sum_{ij} J_{ij} \hat{\sigma}_i^z \hat{\sigma}_j^z.
\end{equation}
We examine a variety of different couplings $J_{ij}$: long-range antiferromagnets $J_{ij} \propto |i - j|^{-\alpha}$ (inspired by current experimental apparatuses~\cite{Koffel, Zhang17,Pagano19}), instances of all-to-all spin glasses having every $J_{ij}$ chosen randomly from $[-1, 1]$, and instances of the MaxCut problem on random 4-regular graphs ($J_{ij}$ being the adjacency matrix of the graph).

All models studied show the same qualitative behavior: the optimal protocol begins and ends with a bang, as it must, and in between has extended annealing portions (possibly punctuated by additional bangs).
We term such protocols ``bang-anneal-bang''.
For concreteness, we shall present the results obtained for the MaxCut problem, with numerics for other models presented in Appendix \ref{app:numerics}.

\begin{figure}[t]
\centering
\includegraphics[width=1.0\columnwidth]{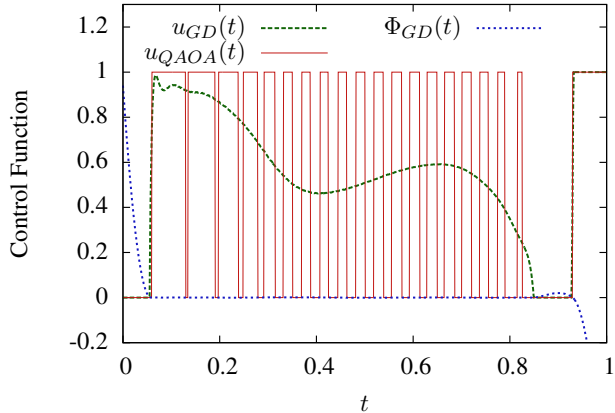}
\caption{Optimal control functions found through either gradient descent ($u_{GD}(t)$) or constrained-time QAOA ($u_{QAOA}(t)$) for a random instance of the MaxCut problem. Also shown is the gradient $\Phi_{GD}(t)$ for the gradient descent method. Parameters: $N = 8$ spins, total time $t_f = 2.0$, $2p = 40$ bangs for the QAOA method.}
\label{fig:grad_descent}
\end{figure}

To find the optimal protocol, we discretize the time evolution in Eqs.~\eqref{eq:optimal_x_equation} and~\eqref{eq:optimal_k_equation}, and apply gradient descent (specifically Nesterov's method \cite{Nesterov}) to $J$ as a functional of $u(t)$.
The order in which the time evolution is performed is important: since $| x(0) \rangle$ is known, we first evolve forward in time to determine $| x(t) \rangle$, then compute $| k(t_f) \rangle$ through Eq.~\eqref{eq:optimal_endpoint_equation}, then evolve \textit{backwards} in time to determine $| k(t) \rangle$.
The gradient descent could become trapped in a false minimum, so we perform multiple trials using different initial choices for $u(t)$.
In practice, false minima appear to be rare, and most initial guesses found the optimum (all protocols shown were found starting from an initial guess of $u(t)=0.5$).
Fig.~\ref{fig:grad_descent} shows a representative example of a protocol thus obtained, denoted $u_{GD}(t)$ (dashed green line), as well as the corresponding $\Phi_{GD}(t)$ (dashed blue line).
As proven, it has bangs at the beginning and end. In the middle, either bangs or smooth anneals are possible based on our theoretical analysis which is part of the reason for numerical study.  Numerically, we always find the middle region dominated by smooth anneals, possibly with an additional bang at the end as seen in Fig.~\ref{fig:grad_descent}.
Also note the consistency between the behavior of $u_{GD}(t)$ and the sign of $\Phi_{GD}(t)$.

\begin{figure}[t]
\centering
\includegraphics[width=1.0\columnwidth]{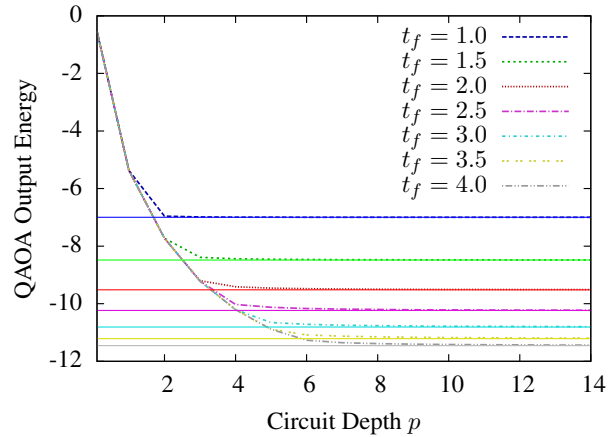}
\caption{QAOA output energy as a function of the circuit depth $p$ (i.e., number of bangs), for a $N = 10$ MaxCut instance on a $4$-regular graph.  Dashed lines are the QAOA energies;  solid horizontal lines are the energies obtained from gradient descent.}
\label{fig:gd_vs_qaoa}
\end{figure}

Fig.~\ref{fig:grad_descent} also shows the result of QAOA for the same instance, using a fixed number of bangs ($p=20$ bangs each of $u = 0$ and $u = 1$) and optimizing over the length of each interval (with the sum constrained to be $t_f$).
The bangs in the middle of the protocol, where gradient descent would produce an annealing segment, are significantly shorter than those at the beginning or end.
This makes sense given the Suzuki-Trotter decomposition:
\begin{equation} \label{eq:Suzuki_Trotter_decomposition}
\begin{aligned}
e^{-i(\beta \hat{B} + \gamma \hat{C})} & = e^{-i \frac{\beta}{2p} \hat{B}} e^{-i \frac{\gamma}{p} \hat{C}} \\
\cdot & \left( e^{-i \frac{\beta}{p} \hat{B}} e^{-i \frac{\gamma}{p} \hat{C}} \right) ^{p-1} e^{-i \frac{\beta}{2p} \hat{B}} + O \left( \frac{1}{p^2} \right) .
\end{aligned}
\end{equation}
A large number of short bangs serves as a reasonable approximation to an annealing segment.
Fig.~\ref{fig:grad_descent} suggests that QAOA is indeed attempting to approximate the bang-anneal-bang protocol found by gradient descent.  Note that this behavior is only seen when QAOA is constrained to a fixed time but with increasing QAOA depth $p$.  QAOA without the time constraint does not approach a Trotterization \cite{Zhou,Pagano19}. 

\begin{figure}[t]
\centering
\includegraphics[width=0.5\textwidth]{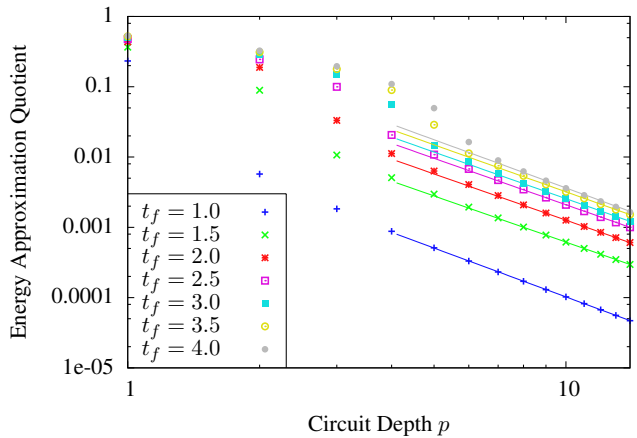}
\caption{Approximation quotient of QAOA output energy versus circuit depth $p$, for the same data as in Fig.~\ref{fig:gd_vs_qaoa}. Solid lines are power-law fits, $Cp^{-\nu}$, to the last 5 data points in each curve. The fitted $\nu \approx 2.2$ for all $t_f$.}
\label{fig:scaling}
\end{figure}

For further evidence, Figs.~\ref{fig:gd_vs_qaoa} and~\ref{fig:scaling} plot respectively the energy and ``approximation quotient'' of the QAOA output state (i.e., $| x(t_f) \rangle$) as functions of the number of bangs $p$.
Here the approximation quotient is defined as $(E_{GD} - E_{QAOA}) / E_{GD}$, where $E_{GD}$ and $E_{QAOA}$ are the output energies of gradient descent and QAOA.
We see that the QAOA protocol performs worse than gradient descent, but with an error that decreases as $p$ increases.
Fitting the error to a power law $C p^{-\nu}$, we find $\nu \approx 2.2$ for all $t_f$.
This is reasonably consistent with the scaling expected from Eq.~\eqref{eq:Suzuki_Trotter_decomposition}.

It should be noted that even though time-constrained QAOA approximates the bang-anneal-bang protocol, it may be the more effective approach in practice. QAOA has a much smaller parameter space to explore -- the durations of the pulses as opposed to an entire function -- and pulses may be simpler to implement experimentally than arbitrary combinations of Hamiltonians.

\textit{Conclusions.}
We have shown that for the control problem of minimizing the energy of a quantum state, the optimal protocol under time constraints is often of the bang-anneal-bang form.
This shows that recent conjectures about the optimality of QAOA based on Pontryagin's principle are not as general as previously thought.  
Our results do not preclude bang-bang entirely, but they do provide the theoretical framework in which non-bang-bang protocols are possible.  Furthermore, our numeric results indicate that these non-bang-bang protocols are extremely common for Ising models; though, there are known non-Ising examples where bang-bang is still optimal \cite{Karzig, Rahmani, Bao}.
Nonetheless, Pontryagin's principle and optimal control theory do serve as valuable tools.
We have used them to prove that the optimal protocol must begin and end with a finite-length bang when not enough time is allowed for the desired state to be reached perfectly.
Furthermore, optimal control theory provides guidance about the form of the protocol in the middle both analytically and numerically through the gradient, $\Phi(t)$.

Keep in mind that these results say nothing of the \textit{practicality} of finding the optimal protocol.
Since the algorithms require simulating the time evolution of the $N$-spin system, they are extremely expensive in computation time and memory on a classical computer.
The main attraction of QAOA is that the time evolution required can be performed on a \textit{quantum} computer, by which we mean simply that the evolution is implemented experimentally on a real system~\cite{Otterbach,Qiang,Pagano19}.
The gradient descent method used in the present paper would be more difficult to implement on the same experimental setups that QAOA is typically implemented on; though other hardware and setups, such as \cite{Novikova} could utilize a similar gradient based method for optimization.
It is obviously of great interest and utility to consider how one might better merge the tools of optimal control theory with current experimental capabilities.

\textit{Acknowledgments.}
We would like to thank P.\ Bienias and Z.\ C.\ Yang for helpful discussions.
The numerics in this paper were done in part using code built upon QAOA code written by S.\ Jordan.
The research of L.T.B. and C.L.B. was supported by National Institute of Standards and Technology (NIST) National Research Council (NRC) Research Postdoctoral Associateship Awards. The research was supported by the U.S. Department of Energy Award Number DE-SC0019449 (time-constrained optimization and analysis) and the U.S. Department of Energy Award Number DE-SC0019139 (quantum variational optimization).

\appendix

\section{Control Problem}
\label{app:control}
The main text uses the conditions for an optimal protocol, derived from optimal control theory \cite{Pontryagin,Kirk}, and in this section we go through the details, deriving Eqs.~(\ref{eq:optimal_x_equation}-\ref{eq:optimal_u_equation}) from the main text.
Our goal in QA/QAOA is to minimize the function
\begin{equation}
        J = \bra{x(t_f)}\hat{C}\ket{x(t_f)},
\end{equation}
subject to the evolution
\begin{equation}
        \ket{\dot{x}(t)} = -i\hat{H}(t)\ket{x(t)},
\end{equation}
\begin{equation}
        \hat{H}(t) = u(t) \hat{B} + (1-u(t))\hat{C}.
\end{equation}
We are subject to a fixed initial state $\ket{x(0)}$.
The function $u(t)$ is our control parameter, bound in the range $u(t)\in[0,1]$.

The following derivation is a reworking of classical optimal control theory \cite{Pontryagin,Kirk} in the context of the QAOA/QA quantum problem.  Classical Optimal Control Theory is just an extension of the Calculus of Variations, tasked now with finding the necessary conditions to optimize $J$.  To ensure that the Schr\"odinger Equation is obeyed in $J$, we introduce the equation as a constraint to be satisfied alongside a Lagrange multiplier $\ket{k(t)}$:
\begin{align}
    \label{eq:Jcomplete}
        J =& \bra{x(t_f)}\hat{C}\ket{x(t_f)}\\\nonumber
        &+\int_{0}^{t_f} dt \left[\bra{k(t)}(-i\hat{H}(t)\ket{x(t)}-\ket{\dot{x}(t)})\right]+\cc.
\end{align}
As per convention, the $\cc$ just refers to the complex conjugate of the preceding equation.

We find the necessary conditions for minimization by looking at what happens to $J$ when we perturb each variable in the same manner as the calculus of variations.  Our goal will be to get $\delta J = 0$, and we are going to need perturbations in $\delta t_f$, $\ket{\delta x_f}$, $\ket{\delta x(t)}$, $\delta u(t)$, and $\ket{\delta k(t)}$.  Each one of these perturbations is independent, and for the relevant ones, we will also need to consider the complex conjugate (handled through the convenient $\cc$ at the end of each necessary line):
\begin{align}
        0=\delta J &= \left[\bra{x(t_f)}\hat{C} - \bra{k(t_f)}\right] \ket{\delta x_f} + \cc\\\nonumber
                & + i\left[\bra{x(t_f)}\hat{H}(t_f)\ket{k(t_f)}\right]\delta t_f+\cc\\\nonumber
                & + \int_{t_0}^{t_f}\!\!dt\! \left[-i\bra{k(t)}\hat{H}(t)+\bra{\dot{k}(t)}\right]\ket{\delta x(t)} +\cc \\\nonumber
                & + \int_{t_0}^{t_f}\!\!dt\! \left[i\bra{x(t)}\pder{\hat{H}}{u}\ket{k(t)}\right]\delta u(t)+\cc \\\nonumber
                & + \int_{t_0}^{t_f}\!\!dt\! \left[-\bra{\dot{x}(t)}+i\bra{x(t)}\hat{H}(t)\right]\ket{\delta k(t)} +\cc.
\end{align}
The third and fifth lines lead to the fairly straightforward equations of motion
\begin{equation}
    \label{eq:app:dotx}
        \ket{\dot{x}^*(t)} = -i \hat{H}^*(t)\ket{x^*(t)},
\end{equation}
\begin{equation}
    \label{eq:app:dotk}
        \ket{\dot{k}^*(t)} = -i \hat{H}^*(t)\ket{k^*(t)}.
\end{equation}
where the $*$ refers to an optimal solution.

Since our final state is allowed to vary, the first line gives us a condition relating $k$ and $x$:
\begin{equation}
        \hat{C}\ket{x^*(t_f)} = \ket{k^*(t_f)}.
\end{equation}
The final condition on time,
\begin{equation}
    \label{eq:app:tcon}
        \left[\bra{x^*(t_f)}\hat{H}^*(t_f)\ket{k^*(t_f)}-\cc\right]\delta t_f=0,
\end{equation}
can either be satisfied automatically if we fix the final time, $\delta t_f=0$, or can give an additional condition if the final time is allowed to vary, $\delta t_f\neq 0$.  In the main text, we assume a fixed $t_f$, so this condition does not appear.

The last condition,
\begin{equation} 
    \left[i\bra{x^*(t)}\pder{\hat{H}}{u}\ket{k(t)}-i\bra{k^*(t)}\pder{\hat{H}}{u}\ket{x^*(t)}\right]\delta u(t)=0,
\end{equation}
is complicated by the fact that $u(t)$ is restricted to the region $u(t)\in[0,1]$.  This restriction means that we need to restrict ourselves down to only  $\delta u(t)$ that are allowed.  This restriction means that the optimum might not be a true extremal point with zero $\frac{\delta J}{\delta u}$ since a disallowed $\delta u$ could still lower $J$.  Pontryagin's Minimum/Maximum Principle \cite{Pontryagin} says that whatever our configuration is needs to be better than what can be achieved through any allowed perturbation of $u(t)$.  Therefore, for a minimum, we need
\begin{equation} 
    \left[i\bra{x^*(t)}\pder{\hat{H}}{u}\ket{k(t)}-i\bra{k^*(t)}\pder{\hat{H}}{u}\ket{x^*(t)}\right]\delta u(t)\geq0,
\end{equation}
for all allowed perturbations $\delta u(t)$.
If we sought a maximum, this would involve $\leq$ instead.

\section{Control Hamiltonian}
\label{app:con_ham}

In this section we will derive and explore the properties of the optimal control Hamiltonian $\mathbb{H}(t)$, presented in Eq.~\ref{eq:control_Hamiltonian}.  Much of this section will be the application of classical Control Theory \cite{Pontryagin, Kirk} and the standard Calculus of Variations to the specific quantum problem at hand.

Consider a general quantity to be minimized (our $J$ or equivalently an action from Lagrangian Mechanics) given by
\begin{equation}
\label{eq:Jgeneral}
J = h(x(t_f),t_f)+\int_{0}^{t_f} dt g(x(t),\dot{x}(t),u(t),t)
\end{equation}
where $x(t)$ is our state variable and $u(t)$ is our control variable, both possibly vectors.  Our goal is to recast this in a traditional Lagrangian format and then perform a Legendre transform to get the corresponding Hamiltonian.

As a first step, note that the ``action'' can be recast as
\begin{equation}
J = \int_{0}^{t_f} dt\left[ g(x(t),\dot{x}(t),u(t),t)+\der{}{t}h(x(t),t)\right]+h(x(0),0),
\end{equation}
where the final $h(x(0),0)$ will be ignored from now on since our initial conditions are fixed and this term will not contribute to the minimization procedure:
\begin{align}
\label{eq:JLagrange}
J &= \int_{0}^{t_f} dt\left[ g(x(t),\dot{x}(t),u(t),t)+\der{}{t}h(x(t),t)\right]\\\nonumber
& = \int_{0}^{t_f} dt\ \mathbb{L}(x(t),\dot{x}(t),u(t),t).
\end{align}

Next, we calculate the generalized momenta via
\begin{equation}
    p(t) = \pder{\mathbb{L}}{\dot{x}},
\end{equation}
and calculate the Hamiltonian
\begin{equation}
    \mathbb{H} = p(t)\cdot\dot{x}(t)-\mathbb{L}.
\end{equation}

To restrict down to our case our state variables are given by $\ket{x(t)}$, $\ket{k(t)}$, $\bra{x(t)}$, $\bra{k(t)}$, and the control Lagrangian can be computed starting from the comparison of the cost function (with a Lagrange multiplier imposing the Schr\"odinger equation):
\begin{align}
    \label{eq:Jcomplete}
        J =& \bra{x(t_f)}\hat{C}\ket{x(t_f)}\\\nonumber
        &+\int_{0}^{t_f} dt \left[\bra{k(t)}(-i\hat{H}(t)\ket{x(t)}-\ket{\dot{x}(t)})\right]+\cc.
\end{align}
with (\ref{eq:Jgeneral}) and continuing through to Eq.~(\ref{eq:JLagrange})
\begin{align}
    \mathbb{L} =& \bra{k(t)}\left(-i\hat{H}(t)\ket{x(t)}-\ket{\dot{x}(t)}\right)+\cc\\\nonumber
    &+\bra{\dot{x}(t)}\hat{C}\ket{x(t)}+\cc.
\end{align}
The generalized momenta are given by
\begin{align}
    \ket{p_x(t)} &= \pder{\mathbb{L}}{\left(\bra{\dot{x}(t)}\right)} = \hat{C}\ket{x(t)}-\ket{k(t)},\\
    \ket{p_k(t)} &= \pder{\mathbb{L}}{\left(\bra{\dot{k}(t)}\right)} = 0,\\
    p_u(t) &= \pder{\mathbb{L}}{\dot{u}(t)} = 0.
\end{align}

Therefore, calculation of the control Hamiltonian gives
\begin{align}
    \label{eq:app:ham}
    \mathbb{H} &= \braket{p_x(t)}{\dot{x}(t)}+\braket{\dot{x}(t)}{p_x(t)}-\mathbb{L}\\\nonumber
    &=i\bra{k(t)}\hat{H}(t)\ket{x(t)}+\cc.
\end{align}

Since $\pder{\mathbb{H}}{t}=0$ (remember that we are treating $u(t)$ as a variable), the control Hamiltonian should be conserved for all time.  To verify this we can take the full time derivative
\begin{align}
    \der{\mathbb{H}}{t} = &-i\bra{\dot{x}(t)}\hat{H}(t)\ket{k(t)}+\cc\\\nonumber
    &-i\bra{x(t)}\hat{H}(t)\ket{\dot{k}(t)}+\cc\\\nonumber
    &-i\bra{x(t)}\dot{u}(t)(\hat{B}-\hat{C})\ket{k(t)} +\cc.
\end{align}
Using the Schr\"odinger equations, 
\begin{align}
    \label{eq:app:dotx}
        \ket{\dot{x}^*(t)} &= -i \hat{H}^*(t)\ket{x^*(t)},\nonumber\\
        \ket{\dot{k}^*(t)} &= -i \hat{H}^*(t)\ket{k^*(t)},
\end{align}
the first two lines exactly cancel with each other.  The leaves the last line which can be rewritten using the notation of the main text as
\begin{align}
    \der{\mathbb{H}}{t} &= -i\bra{x(t)}\dot{u}(t)(\hat{B}-\hat{C})\ket{k(t)} +\cc\\\nonumber
    &=-\dot{u}(t)\Phi(t).
\end{align}
For optimal protocols we either are in a bang region where $u(t)$ is constant implying $\dot{u}(t)=0$ or we are in a singularity where $\Phi(t)=0$.  Therefore, $\der{\mathbb{H}}{t}=0$ and the control Hamiltonian must be a conserved quantity.

Finally, we consider what modifications would need to be done to this picture by including a soft time constraint.  Using the primes to refer to the setting where we have a soft time constraint and $t_f$ is allowed to vary and unprimed quantities to refer to the original $t_f$ fixed problem, we can relate
\begin{equation}
    J' = J+ \lambda t_f
\end{equation}
where $\lambda$ gauges the strength of the soft constraint and makes additional time usage unfavorable.  Going through the derivation, we get the following modifications
\begin{align}
    \mathbb{L}' &= \mathbb{L} + \lambda,\\
    \mathbb{H}' &= \mathbb{H}-\lambda.
\end{align}
obviously since $\lambda$ is time independent, $\mathbb{H}'$ is still a conserved quantity.

\section{Soft Time Constraints}
\label{app:soft_con}
\begin{figure}
        \begin{center}
                \includegraphics[width=0.5\textwidth]{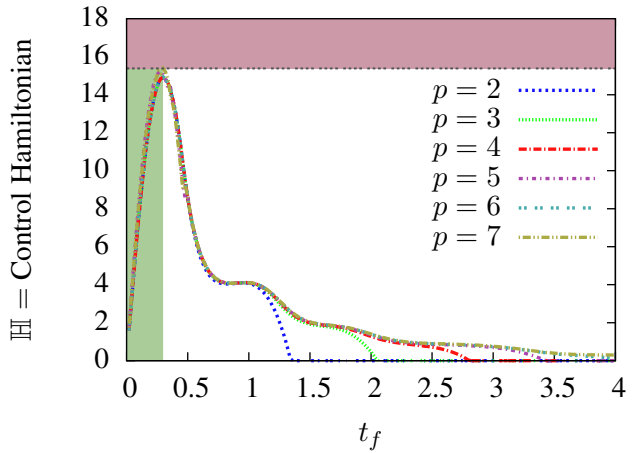}
        \end{center}
        \caption{Optimal control Hamiltonian, $\mathbb{H}(t)$ as a function of the allowed time, $t_f$. Here we consider  hard-time constrained QAOA for a variety of $p$ values, all for the same $n=10$ instance of MaxCut on a random (representative) $4$-regular graph.  The time at which $\mathbb{H}$ turns to zero is the time at which the procedure has more time than unconstrained QAOA would use for that value of $p$.  If a soft-constraint, $\lambda$, is applied, the resulting time for the optimal procedure can be extracted by finding the $t_f$ that results in $\lambda = \mathbb{H}(t)$ (for the largest $t_f$).  This means that no soft-constrained QAOA protocols will have times in the shaded green region, and there are no $t_f\neq0$ protocols for $\lambda$ in the shaded red region or above.}
        \label{fig:con_H}
\end{figure}
This section explores the different forms of soft and hard time constraints, providing the details and derivations behind the explanation of ``\textit{Time constraints}'' in the main text.  This derivation also highlights the physical meaning of the control Hamiltonian, $\mathbb{H}(t)$, discussed in the previous section \ref{app:con_ham}.

We will consider two forms of time constraint: the first is a hard-constraint where we just give the system time $t_f$ as considered in the main text, and the second imposes a linear cost for each additional amount of time taken.  In terms of the hard constraint cost function (unprimed), the soft constraint cost function (primed) is
\begin{equation}
        \label{eq:J_soft}
        J' = \bra{x(t_f)}\hat{C}\ket{x(t_f)}+\lambda t_f = J+\lambda t_f.
\end{equation}
For the most part, this does not modify the equations of motion except the condition due to variations in the final time:
\begin{equation}
    \label{eq:app:tcon}
        \left[\bra{x^*(t_f)}\hat{H}^*(t_f)\ket{k^*(t_f)}-\cc\right]\delta t_f=0,
\end{equation}
which does not apply in the hard constraint case since $\delta t_f = 0$ and becomes
\begin{equation}
        \label{eq:bcend2_soft}
        (\bra{x^*(t_f)}\hat{H}^*(t_f)\ket{k^*(t_f)}-\cc) = i\lambda,
\end{equation}
 for the soft constraints which can be thought of as a condition solely on $t_f$.

When $\lambda=0$ and $t_f$ is allowed to be free, Eq.~(\ref{eq:app:tcon}) combined with the form of $\mathbb{H}(t_f)$, Eq.~(\ref{eq:app:ham}), implies that $\mathbb{H}(t)=0$, and we can verify this numerically in both the quantum adiabatic limit and for QAOA protocols that are constrained by $p$ but not $t_f$.

However, the introduction of $\lambda \neq 0$ modifies the $\lambda = 0$ control Hamiltonian $\mathbb{H}(t)$ so that the soft-constrained version (primed) is
\begin{equation}
    \mathbb{H}'(t) = \mathbb{H}(t)-\lambda.
\end{equation}
Since this problem has a free final time, $\mathbb{H}'(t) = 0$ (see Eq.~(\ref{eq:bcend2_soft})), which means that the original control Hamiltonian for an equivalent hard constraint problem has $\mathbb{H}(t) = \lambda$.

Furthermore, if we look at a hard-constrained problem with a cutoff of $t_f$, the equations of motion are again the same, but the control Hamiltonian is equal to a non-zero constant, with the constant being dependent on $t_f$.  Since the equations of motion are identical, this constant is just $\lambda$ again and just dictates how much soft-constraint would lead to the same $t_f$ as our hard-constrained solution.  In practice, the dependence of $\lambda$ on $t_f$ (or vice-versa) is heavily influenced by the nature of the problem and how quickly the system approaches its ground state energy.

In Fig.~(\ref{fig:con_H}) we plot the value of the control Hamiltonian versus $t_f$ for $t_f$-constrained QAOA applied to a random MaxCut instance.  To read off the equivalence between soft and hard constraints in this plot, draw a horizontal line from the $y$ axis at the value of $\lambda$ you have.  Where that line intersects the desired $p$ curve is the time that a soft-constrained problem with $\lambda$ would prefer.  Note that the $\lambda$ line could intersect at multiple points, in this case, take the right-most point.  For instance the intersections in the green shaded region will never be preferred by soft-constrained QAOA.  Furthermore, any soft constraints, $\lambda$, above the black dashed line are so constricted that they prefer a $t_f=0$ solution with the $p$ shown.  As $p$ increases further, the curves slowly start encroaching upward into this region.

Since soft and hard constraints are theoretically equivalent through this factor $\lambda$, we focus on hard time constraints for most of the paper.
Also note that $\mathbb{H}(t_f)=\lambda = i\bra{x(t_f)}\comm{\hat{C}}{\hat{B}}\ket{x(t_f)}$ can be measured and could be used as an estimate of the error rates in the system after a QAOA variational loop has been completed.

\section{Non-singular Bang Length}
\label{app:bang_len}
One of our key results is that all optimal protocols must begin with a finite length bang (assuming we start in an eigenstate of $\hat{B}$) and end with a finite length bang (assuming our cost function is an expecation value of $\hat{C}$).  One important question is whether we can estimate how long these respective bangs should be.  Numerically, as seen in Fig.~(\ref{fig:gd_vs_qaoa}), these initial and final bangs remain large on the timescale of the system.

While the initial and final bangs are undergoing Hamiltonian evolution under a static Hamiltonian, it is not possible to use simple a priori arguments to determine the exact lengths of the bangs.  This is because of a lack of boundary information.  For instance, in the initial bang, we know the starting point of $\ket{x(0)}$ but not the starting point of $\ket{k(0)}$, both of which we need to determine the first value of $t$ such that $\Phi(t)=0$ which would herald the probable end of the bang.  Similarly, at the end, we know that $\ket{k(t_f)} = \hat{C}\ket{x(t_f)}$, but there is no a priori way of knowing what $\ket{x(t_f)}$ is, that being the goal of these algorithms.

For both the initial and final bang, the length of the bang is determined by the time it takes $|\Phi(t)|$ to transition from $\lambda$ to $0$.  Therefore, the value of $\lambda$ will determine the lion's share of how long the bang takes.  As described in the main text, $\lambda$ can be thought of as the penalty to the cost function for each extra amount of calculation.  Therefore, the function $\lambda(t_f)$ will depend primarily on how the achieved QAOA energy scales with $t_f$.  For instance, if the QAOA energy scales with $\mathcal{O}(1/t_f)$ towards the true ground state, then the cost function $J'$ in Eq.~(\ref{eq:J_soft}) will be the trade-off between $1/t_f$ scaling in the energy and $t_f$ scaling in the time cost, resulting in some balance that produces a preferred $t_f$ curve like that seen in Fig.~\ref{fig:con_H}.  We expect $\lambda$ to decrease with increasing $t_f$, but the exact form of that decrease will be problem specific as discussed in the main text.

In addition to $\lambda$, one of the key factors determining the length of the bangs is how fast $\Phi(t)$ can change.  To first order, we can approximate this by $\dot{\Phi}(t_f)$ and $\dot{\Phi}(0)$.  At the final time, it is easy to see that
\begin{equation}
    \dot{\Phi}(t_f) = \bra{x(t_f)}\comm{\comm{\hat{B}}{\hat{C}}}{\hat{C}}\ket{x(t_f)}.
\end{equation}
For large enough $t_f$, this quantity should depend primarily on the nature of the ground state and a few excited states.  Because of our boundary conditions, we cannot write $\dot{\Phi}(0)$ in terms of only $\ket{x(0)}$ and not $\ket{k(0)}$, but we similarly expect (and see numerically) that this quantity is roughly constant.

Therefore, up to multiplicative constants, we expect $\lambda$ to determine the scaling of the sizes of the bangs.  Therefore, these bangs should become smaller and smaller as $t_f$ is increased. Eventually in the true $t_f\to\infty$ adiabatic limit, these bangs disappear recovering the standard form expected for quantum adiabatic computing.

\section{Classifying Singularities}
\label{app:singularities}
In the main text we prove that all optimal protocols begin and end with non-singular bang regions.  In this section, we find no evidence that singular regions (i.e. smooth annealing regions) cannot exist frequently in the middle region, and we derive the conditions for such singularities.  In fact numerically as seen in Fig.~(\ref{fig:gd_vs_qaoa}), we find that such singular regions are common in the true optimal protocol.

First, a singular region implies that $\Phi(t)=0$ for an extended period of time, which by the constancy of the optimal control Hamiltonian [see Eq.~(\ref{eq:control_Hamiltonian}) or Eq.~(\ref{eq:app:ham})] means that in a singular region with $u^*(t)\in(0,1)$, $\Phi_C(t)=\lambda$ and $\Phi_B(t)=\lambda$.

In order for $\Phi_C(t)$ and $\Phi_B(t)$ to remain constant, all their time derivatives must be zero.  Simple differentiation and application of the Shr\"odinger equation show that the first derivative condition reduces to
\begin{equation}
     \Phi_{\comm{\hat{B}}{\hat{C}}}=0,
\end{equation}
where $\Phi_X$ for any operator $X$ is defined by
\begin{equation} \label{eq:phi_X_definition}
\Phi_X(t) \equiv i \langle k(t) | \hat{X} | x(t) \rangle + \textrm{c.c},
\end{equation}

The second derivatives give the condition
\begin{align}
 0 =\ & \Phi_{\comm{\comm{\hat{B}}{\hat{C}}}{\hat{B}}}u^*(t) + \Phi_{\comm{\comm{\hat{B}}{\hat{C}}}{\hat{C}}}(1-u^*(t)).
\end{align}

At this level, there are three possibilities.  
The first possibility is that 
\begin{equation}
     \label{eq:u_sing}
     u^*(t) = \frac{\Phi_{\comm{\comm{\hat{B}}{\hat{C}}}{\hat{C}}}(t)}{\Phi_{\comm{\comm{\hat{B}}{\hat{C}}}{\hat{B}}}(t)-\Phi_{\comm{\comm{\hat{B}}{\hat{C}}}{\hat{C}}}(t)}.
\end{equation}
A similar form of singular control was explored geometrically in \cite{Lin} where it was used to show a singular control for the Grover search problem that outperforms the traditional Grover problem (but has the same asymptotic scaling).

The second possibility is that only  $\Phi_{\comm{\comm{\hat{B}}{\hat{C}}}{\hat{C}}}(t)=0$, but this necessitates $u^*(t)=0$, and while this is a singular protocol, it matches the form of a bang-bang procedure.  There are known examples (for non-Ising models) of cases where such a bang-bang singularity is optimal \cite{Bao}.

The last would be that $\Phi_{\comm{\comm{\hat{B}}{\hat{C}}}{\hat{B}}}(t)=0$ and $\Phi_{\comm{\comm{\hat{B}}{\hat{C}}}{\hat{C}}}(t)=0$ which would necessitate considering higher derivatives, and this eventuality is discussed later.  

If we have the first case, that just guarantees the derivatives are zero at this level, so we would need to go up to higher levels and verify compliance there.  That compliance could be through a condition on $u(t)$ such as seen in Eq.~(\ref{eq:u_sing}), but as we go up the ladder to higher derivatives, the requirements become overconstrained with two equations to satisfy at the next level, four at the next, and so on, making this possibility unlikely.

Another possibility is that all the $\Phi_X(t)=0$ due to the nature of the states $\ket{x(t)}$ and $\ket{k(t)}$ which is the only possible source of singularties in classical systems linear in $\vec{x}(t)$ and $u(t)$ and then only if the control system lacks full controllability.  This is not possible in our case by construction because our initial state, as a ground state of $\hat{B}$ must lie within the subspace defined by the shared symmetries of $\hat{B}$ and $\hat{C}$.  Therefore, $\ket{x(t)}$ (and also $\ket{k(t)}$) must lie entirely within this same symmetry subspace.  The Lie algebra generated by $\hat{B}$ and $\hat{C}$ can move us around fully within that subspace, with symmetries being what restricts down the controllability of the system, meaning that some operator within the Lie algebra generated by nested commutators of $\hat{B}$ and $\hat{C}$ must have a non-zero matrix element between $\ket{x(t)}$ and $\ket{k(t)}$.

For the reasons discussed above, we expect most singularities to be of the form shown in Eq.~(\ref{eq:u_sing}) with other forms increasingly unlikely.  The big question is whether we will see this form of singularity or a bang-bang form in general as the global, rather than local, optimum.

These singularities are similar to the forms seen in more traditional quantum optimal control literature for their version of singularities \cite{Riviello,Russell}.

 It should be noted that all the conditions listed here are necessary conditions for optimality, not sufficient.  In order for these conditions to be sufficient, $J$ would need to be a convex function of $\ket{x(t)}$ \cite{Mangasarian} which is not true in general for quantum systems.  In quantum optimal control problems with no constraints on the control parameter and other conditions such as full controllability, false minima or traps that satisfy the optimal conditions but are not true optima are exceedingly rare \cite{Russell}. However, when control and time constraints apply (as in our case), traps become more likely.

Our case does have a large number of constraints, so the standard logic from quantum optimal control theory about the non-idealness of singular solutions \cite{Riviello,Russell} does not hold here.  As was seen in the main text and discussed further in the next part of this section, in Ising models at least, the singularity from Eq.~(\ref{eq:u_sing}) turns out to be the global minimum for the control landscape.

\subsection{Forms of Singularities in Practice}
\begin{figure}
        \begin{center}
                \includegraphics[width=0.5\textwidth]{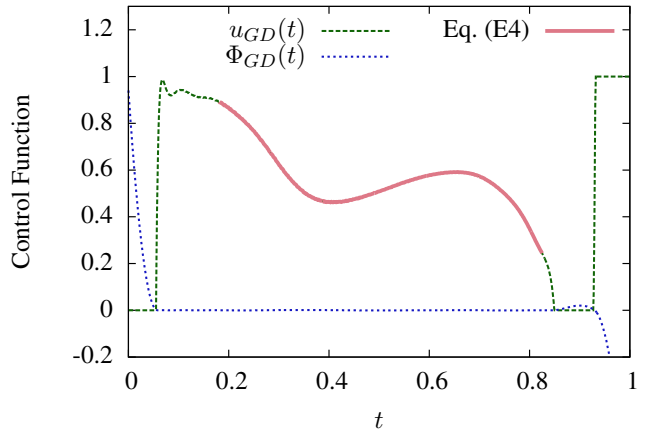}
        \end{center}
        \caption{This plot is mostly a reproduction of Fig.~(\ref{fig:gd_vs_qaoa}).  The additional solid red highlighting shows the region where $u(t)$ obeys the singular condition outlined in Eq.~(\ref{eq:u_sing}).  In other regions, the control function is presumably obeying a higher order singularity condition, but some of the discrepancy could be due to numerical errors.}
        \label{fig:grad_using}
\end{figure}

While Fig.~(\ref{fig:gd_vs_qaoa}) was constructed using a gradient descent method, the singular regions must still follow the different forms of singularities possible for this problem as discussed in the previous section.  This section discusses how the analytic results of the previous section apply to our numerics.

For instance, a large question is whether $u(t)$ in Fig.~(\ref{fig:gd_vs_qaoa}) obeys the singularity condition in Eq.~(\ref{eq:u_sing}) or if it obeys some other condition derived from a higher derivative as discussed in the previous section.  The answer to this question is twofold.  For the most part, the singular $u(t)$ does obey Eq.~(\ref{eq:u_sing}), especially in the middle of the smooth curve.  However, it does not always follow this condition.

In Fig.~\ref{fig:grad_using}, we reproduce Fig.~(\ref{fig:gd_vs_qaoa}), highlighting the region where the singular control obeys Eq.~(\ref{eq:u_sing}) to within a numerical tolerance.  Notice that the highlighted region does not encompass the entire singular region.  This implies that for portions of the evolution, $u(t)$ is obeying some higher order singularity condition as discussed, but not derived, in the main text.

Note that our simulation requires a discretization of $u(t)$ into 1001 $\Delta t$ steps and that our gradient descent only achieves the optimal up to some precision based on how long we run it.  Therefore, the $u(t)$ shown in Fig.~\ref{fig:grad_using} could have numerical errors shortening the range of agreement with Eq.~(\ref{eq:u_sing}), but it is possible that higher derivative conditions are relevant here.

\section{Additional Numerics} \label{app:numerics}

\begin{figure}
        \begin{center}
                \includegraphics[width=0.5\textwidth]{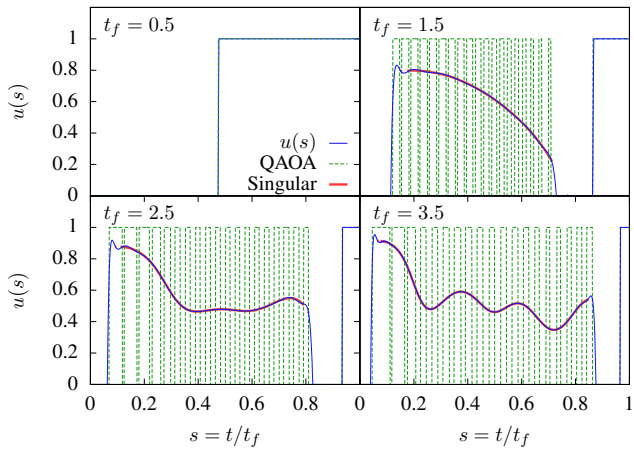}
        \end{center}
        \caption{This plot shows results for a particular instance of our Random Ising (RI) model, showing the optimal $u(t)$ protocol, a $p=20$ constrained QAOA protocol that takes the same amount of time, and a highlighting of where the $u(s)$ matches the singular condition from Eq.~\ref{eq:u_sing}.  Note that the singular condition does not match for the entire annealing region, and we claim that this deviation is due mostly to premature stopping of the gradient descent procedure.  Each plot represents a different final $t_f$ value.}
        \label{fig:RI}
\end{figure}

\begin{figure}
        \begin{center}
                \includegraphics[width=0.5\textwidth]{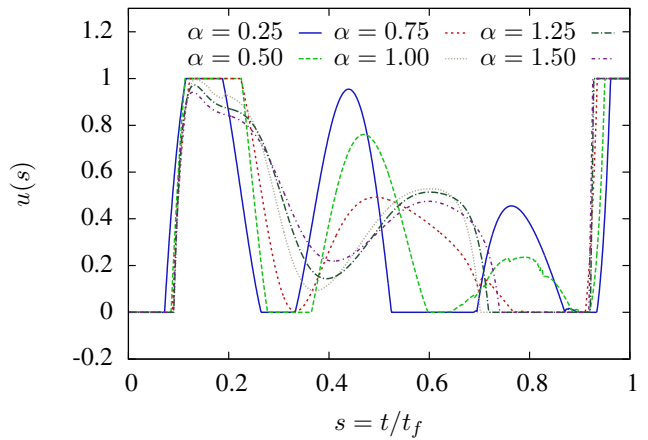}
        \end{center}
        \caption{This plot shows results for the Long Range Ising model (LR) for various values of $\alpha$ at $n=5$ and $t_f=2$, showing the optimal $u(t)$ protocol.  The curves still roughly follow a bang-anneal-bang pattern, but for some of the plots, the annealing region's oscillations are large enough to flatten into bangs at the peaks.}
        \label{fig:LR5}
\end{figure}

\begin{figure}
        \begin{center}
                \includegraphics[width=0.5\textwidth]{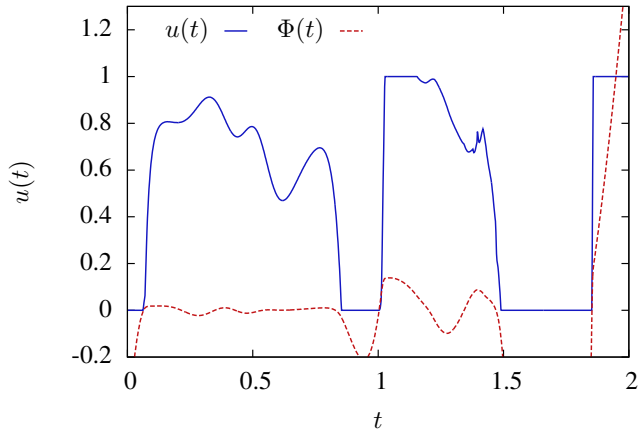}
        \end{center}
        \caption{This plot shows the optimal protocol for the Heisenberg model where the problem Hamiltonian is given by $\hat{C}=\sum_{i<j} \frac{J_{ij}}{3} \left(\hat{\sigma}_i^x \hat{\sigma}_j^x+\hat{\sigma}_i^y \hat{\sigma}_j^y+\hat{\sigma}_i^z \hat{\sigma}_j^z\right)$.  The specific $J_{ij}$ matrix here is the same one used in Fig. (1) of the main text.  Note that the $\Phi(t)$ indicates that the gradient descent done to produce this figure was not as thorough.  This is due largely to numerical limits and greater complexity of simulating the given $\hat{C}$ relative to the diagonal Ising model.  Nevertheless, this figure demonstrates clear initial and final bangs with the interior region exhibiting both annealing and bang-bang behavior as our analytics predict.}
        \label{fig:mixed}
\end{figure}

In the main text we focus on a particular instance of the MaxCut problem to provide example numerics.  In this section, we expand to examining other problems, most Ising but with also some Heisenberg models, all of which exhibit roughly the same qualitative behavior.  We also examine more in depth the behavior of these optimal bang-anneal-bang curves as $t_f$ and $n$ change.  There are two additional Ising models that we will be considering here, both conforming to the basic form
\begin{equation}
        \hat{C} = \sum_{i<j} J_{ij} \hat{\sigma}_i^z \hat{\sigma}_j^z.
\end{equation}
The first model, Randomized Ising (abbreviated RI), will just involve a randomized all-to-all connected $J_{ij}$ matrix where each coupling is chosen randomly from the range $[-1,1]$.  The second model, Long-Range Ising (abbreviated LR), is motivated by our experimental collaboration \cite{Pagano19} and involves a 1D chain of qubits interacting via
\begin{equation}
        J_{ij} = \frac{J}{|i-j|^\alpha},
\end{equation}
where we have taken $J=1$ in our dimensionless calculations.

First in Fig.~\ref{fig:RI} we show data from an instance of our RI model.  In blue, we show the optimal $u(t/t_f)$ procedure; in green we show a $p=20$ constrained QAOA procedure for the same problem; and in red we show where the protocol matches up with the singular condition from Eq.~\ref{eq:u_sing}.  The singular condition does have gaps which we attribute mostly to the numerical precision of our gradient descent method.  Note that this figure also demonstrates some of the key features we see in these protocols.  The initial and final bangs become shorter and shorter as $t_f$ increases, in accordance with our predictions from Section \ref{app:bang_len}.  Finally, the annealing region gains more structure and the number of oscillations grows as $t_f$ increases.  The structure of this oscillatory pattern is being explored further in follow-up work including a subset of the current authors \cite{curves_paper}.

In Fig.~\ref{fig:LR5} we show results from the LR model for $n=5$ and $t_f=2$.  Notice that the oscillations in the middle of the evolution  have large enough amplitude that they reach the edge of the allowed domain and become bangs, at least for low $\alpha$ plots.  It is also interesting that for $\alpha\geq1$, the procedures are very similar in appearance.  In general, we have noticed that similar problem instances often lead to qualitatively similar curves.

Finally to give a non-Ising example, we present Fig.~\ref{fig:mixed} which shows the optimal protocol when the Hamiltonian corresponds to a Heisenberg model:
\begin{equation}
    \hat{C}=\sum_{i<j} \frac{J_{ij}}{3} \left(\hat{\sigma}_i^x \hat{\sigma}_j^x+\hat{\sigma}_i^y \hat{\sigma}_j^y+\hat{\sigma}_i^z \hat{\sigma}_j^z\right).
\end{equation}
The specific $J_{ij}$ matrix for this figure is the connectivity graph for an $n=8$ 4-regular graph (the same one used in Fig. (1) of the main text).  Most notably, this optimal protocol still maintains the bangs at the beginning and end and strongly requires them based off the sign and magnitude of $\Phi(t)$.  In the interior region, both bangs and anneals are seen.  The $\Phi(t)$ here is rougher and indicates that further fine-tuning with gradient descent is possible.  This tuning is much harder numerically for this model since the current $\hat{C}$ is non-diagonal and more complicated to simulate.  Overall, this plot does match with the results of our analytic analysis.


\begin{thebibliography}{99}

\bibitem{Farhi2000} E. Farhi, J. Goldstone, S. Gutmann, M. Sipser, arXiv:0001106 (2000).

\bibitem{Kadowaki} T. Kadowaki, H. Nishimori, \textit{Phys. Rev. E} {\bf 58}, 5355 (1998).

\bibitem{Farhi2014} E. Farhi, J. Goldstone, S. Gutman, 	arXiv:1411.4028 (2014).

\bibitem{Yang} Z. C. Yang, A. Rahmani, A. Shabani, H. Neven, C. Chamon, \textit{Phys. Rev. X}, {\bf 7}), 021027 (2017).

\bibitem{Bapat2018} A. Bapat, S. Jordan, arXiv:1812.02746 (2018).

\bibitem{Mbeng} G. B. Mbeng, R. Fazio, G. Santoro, arXiv:1906.08948 (2019).

\bibitem{Lin} C. Lin, Y. Wang, G. Kolesov, U. Kalabi\'c, \textit{Phys. Rev. A} {\bf 100}, 022327 (2019).

\bibitem{Pontryagin} L. S. Pontryagin, V. G. Boltyanskii, R. V. Gamkrelidze, E. F. Mishchenko, \textit{The Mathematical Theory of Optimal Processes}, (Interscience Publishers, New York, 1962).

\bibitem{Jansen} S. Jansen, M. Ruskai, R. Seiler, 	\textit{J. Math. Phys.} {\bf 48}, 102111 (2007).

\bibitem{Roland} J. Roland, N. J. Cerf, \textit{Phys. Rev. A} {\bf 65}, 042308 (2002).

\bibitem{Jarret} M. Jarret, B. Lackey, A. Liu, K. Wan, arXiv:1810.04686 (2018).

\bibitem{Farhi2016} E. Farhi, A. Harrow, arXiv:1602.07674 (2016).

\bibitem{Aharonov} D. Aharonov, W. van Dam, J. Kempe, Z. Landau, S. Lloyd, O. Regev, Proc. 45th FOCS, p. 42-51 (2004).

\bibitem{Lloyd} S. Lloyd, arXiv:1812.11075 (2018).

\bibitem{Rabitz2000} H. Rabitz, R. de Vivie-Riedle, M. Motzkus, K. Kompa, \textit{Science}, Vol. 288, 5467, pp. 824-828 (2000).

\bibitem{Werschnik} J. Werschnik, E. K. U. Gross, \textit{Journal of Phys. B}, Vol. 40, 18, pp. R175--R211 (2007).

\bibitem{Brif11} C. Brif,  R. Chakrabarti,  H. Rabitz, \textit{New Journal of Phys.}, Vol. 12 (2010).


\bibitem{Palao02} J. P. Palao, R. Kosloff, \textit{Phys. Rev. Lett.} {\bf 89}, 188301 (2002).

\bibitem{Palao03} J. P. Palao, R. Kosloff, \textit{Phys. Rev. A} {\bf 68}, 062308 (2003).

\bibitem{Khaneja} N. Khaneja, T. Reiss, C. Kehlet, T. Schulte-Herbrüggen, S. J. Glaser, \textit{Journal of Magnetic Resonance} 172, pp. 296-305 (2005).

\bibitem{Montangero} S. Montangero, T. Calarco, R. Fazio, \textit{Phys. Rev. Lett.} {\bf 99}, 170501 (2007).


\bibitem{Gorshkov} A. V. Gorshkov, T. Calarco, M. D. Lukin, A. S. S{\o}rensen, \textit{Phys. Rev. A} {\bf 77}, 043806 (2008).

\bibitem{Grace} M. Grace, C. Brif, H. Rabitz, I. A. Walmsley, R. L. Kosut, D. A. Lidar, \textit{Journal of Physics B} {\bf40}, 9 (2009).

\bibitem{Cui} J. Cui, R. van Bijnen, T. Pohl, S. Montangero, T. Calarco, \textit{Quantum Sci. Technol.} {\bf2}  035006 (2017).

\bibitem{Omran} A. Omran, H. Levine, A. Keesling, G. Semeghini, T. T. Wang, S. Ebadi, H. Bernien, A. S. Zibrov, H. Pichler, S. Choi, J. Cui, M. Rossignolo, P. Rembold, S. Montangero, T. Calarco, M. Endres, M. Greiner, V. Vuleti\'c, M. D. Lukin, \textit{Science}, Vol. 365, 6453, pp. 570-574 (2019).

\bibitem{Riviello15} G. Riviello, K. M. Tibbetts, C. Brif, R. Long, R. Wu, T. Ho, H. Rabitz, \textit{Phys. Rev. A} {\bf 91}, 043401 (2015).

\bibitem{Koffel} T. Koffel, M. Lewenstein, L. Tagliacozzo, \textit{Phys. Rev. Lett.} {\bf 109}, 267203 (2012).

\bibitem{Zhang17} J. Zhang, G. Pagano, P. W. Hess, A. Kyprianidis, P. Becker, H. Kaplan, A. V. Gorshkov, Z. X. Gong, C. Monroe, \textit{Nature} {\bf 551}, 601–604 (2017).

\bibitem{Pagano19} G. Pagano, A. Bapat, P. Becker, K. S. Collins, A. De, P. W. Hess, H. B. Kaplan, A. Kyprianidis, W. L. Tan, C. Baldwin, L. T. Brady, A. Deshpande, F. Liu, S. Jordan, A. V. Gorshkov, C. Monroe, arXiv:1906.02700 (2019).


\bibitem{Mangasarian} O. L. Mangasarian, \textit{J. SIAM Control}, Vol.\ 4, No.\ 1 (1966).

\bibitem{Wu} R. Wu, R. Long, J. Dominy, T. Ho, H. Rabitz, \textit{Phys. Rev. A} {\bf 86}, 013405 (2012).

\bibitem{Riviello} G. Riviello, C. Brif, R. Long, R. Wu, K. M. Tibbetts, T. Ho, H. Rabitz, \textit{Phys. Rev. A} {\bf 90}, 013404 (2014).

\bibitem{Russell} B. Russell, H. Rabitz, R. Wu, arXiv:1608.06198 (2016).

\bibitem{Kirk} D. E. Kirk, \textit{Optimal Control Theory: An Introduction}, Dover Publications (1998).

\bibitem{Nesterov} Y. Nesterov, \textit{Introductory Lectures on Convex Optimization}, Springer (2004).

\bibitem{Novikova} I. Novikova, A. V. Gorshkov, D. F. Phillips, A. S. S{\o}rensen, M. D. Lukin, R. L. Walsworth, \textit{Phys. Rev. Lett.} {\bf 98}, 243602 (2007).


\bibitem{Zhou} L. Zhou, S. Wang, S. Choi, H. Pichler, M. D. Lukin, arXiv:1812.01041 (2018).

\bibitem{Rahmani} A. Rahmani, T. Kitagawa, E. Demler, C. Chamon, \textit{Phys. Rev. A} {\bf 87}, 043607 (2013).

\bibitem{Karzig} T. Karzig, A. Rahmani, F. von Oppen, G. Refael, \textit{Phys. Rev. B} {\bf 91}, 201404(R) (2015).

\bibitem{Bao} S. Bao, S. Kleer, R. Wang, A. Rahmani, \textit{Phys. Rev. A} {\bf 97}, 062343 (2018).


\bibitem{Qiang} X. Qiang, X. Zhou, J. Wang, C. M. Wilkes, T. Loke,
S. O’Gara, L. Kling, G. D. Marshall, R. Santagati, T. C.
Ralph, J. B. Wang, J. L. O’Brien, M. G. Thompson,
J. C. F. Matthews, Nat. Photonics {\bf 12}, 534 (2018).

\bibitem{Otterbach} J. S. Otterbach, R. Manenti, N. Alidoust, A. Bestwick,
M. Block, B. Bloom, S. Caldwell, N. Didier, E. S. Fried,
S. Hong, P. Karalekas, C. B. Osborn, A. Papageorge,
E. C. Peterson, G. Prawiroatmodjo, N. Rubin, C. A.
Ryan, D. Scarabelli, M. Scheer, E. A. Sete, P. Sivarajah,
R. S. Smith, A. Staley, N. Tezak, W. J. Zeng, A. Hudson, B. R. Johnson, M. Reagor, M. P. da Silva,
C. Rigetti,  arXiv:1712.05771 (2017).

\bibitem{curves_paper} L. T. Brady, L. Kocia, P. Bienias, A. Bapat, Y. Kharkov, A. V. Gorshkov, \textit{in preperation}.


\end{thebibliography}
\end{document}